\documentclass[fleqn]{2023SCGE}
\usepackage{hyperref}

\def\bl#1\el{\begin{align}#1\end{align}}
\def\be#1\ee{\begin{equation}#1\end{equation}}
\def\nn{\nonumber}

\begin{document}

\ensubject{subject}

%%%%%%%%%%%%%%%%%%%%%%%%%%%%%%%%%%%%%%%%%%%%%%%%%%%%%%%
%%% Authors do not modify the information below
%%% ????????????????
%%% ??????????, ????????????{}, ???????????????????
%Letter to the Editor??Article%??????
\ArticleType{Article}%??Article
%\SpecialTopic{}%???????
\Year{2025}
\Month{}
\Vol{68}
\No{4}
\DOI{10.1007/s11433-024-2573-3}
\ArtNo{249512}
\ReceiveDate{October 14, 2024}
\AcceptDate{December 17, 2024}
\OnlineDate{February 19, 2025}
%%%%%%%%%%%%%%%%%%%%%%%%%%%%%%%%%%%%%%%%%%%%%%%%%%%%%%%

%%% title: ????
%%%   \title{title}{title for citation}
\title{Space-based optical lattice clocks as gravitational wave detectors in search for new physics}{Space-based optical lattice clocks as gravitational wave detectors in search for new physics}

%%%   \author[number]{Full name}{{email@xxx.com}}

\author[1,2]{Bo Wang}{{ymwangbo@ustc.edu.cn }}
\author[1,2]{Bichu Li}{}
\author[1,2]{Qianqian Xiao}{}
\author[1,2]{Geyu Mo}{}
\author[1,2,3]{Yi-Fu Cai}{{yifucai@ustc.edu.cn}}

\AuthorMark{B. Wang}

\AuthorCitation{B. Wang, B. Li, Q. Xiao, G. Mo, Y.-F. Cai}

\address[1]{Department of Astronomy, School of Physical Sciences, \\University of Science and Technology of China, 96 Jinzhai Road, Hefei, Anhui 230026, China}
\address[2]{CAS Key Laboratory for Research in Galaxies and Cosmology, School of Astronomy and Space Science, \\University of Science and Technology of China, 96 Jinzhai Road, Hefei, Anhui 230026, China;}
\address[3]{Deep Space Exploration Laboratory, Hefei 230088, China}

%\contributions{}%????????

%%% Abstract. ??
\abstract{We investigate the sensitivity and performance of space-based Optical Lattice Clocks (OLCs) in detecting gravitational waves, in particular the Stochastic Gravitational Wave Background (SGWB) at low frequencies $(10^{-4}, 1) \rm Hz$, which are inaccessible to ground-based detectors. 
We first analyze the response characteristics of a single OLC detector for SGWB detection and compare its sensitivity with that of Laser Interferometer Space Antenna (LISA). 
Due to longer arm lengths, space-based OLC detectors can exhibit unique frequency responses and enhance the capability to detect SGWB in the low-frequency range, but the sensitivity of a single OLC detector remains insufficient overall compared to LISA. 
Then, as a preliminary plan, we adopt a method of cross-correlation on two OLC detectors to improve the signal-to-noise ratio (SNR).
This method leverages the uncorrelated origins but statistically similar properties of noise in two detectors while the SGWB signal is correlated between them, thus achieving effective noise suppression and sensitivity enhancement. 
Future advancements in OLC stability are expected to further enhance their detection performance.
This work highlights the potential of OLC detectors as a promising platform for SGWB detection, offering complementary capabilities to LISA, and opening an observational window into more astrophysical sources and the early universe.}%ÕªÒª

%%% Keywords. ?????


\maketitle

%\tableofcontents%?????

%%%%%%%%%%%%%%%%%%%%%%%%%%%%%%%%%%%%%%%%%%%%%%%%%%%%%%%
%%% The main text. ???????
%???????????????????\cref{fig1}
%\twocolumn\onecolumn
%%%%%%%%%%%%%%%%%%%%%%%%%%%%%%%%%%%%%%%%%%%%%%%%%%%%%%%
\begin{multicols}{2}
\section{Introduction}
\label{sec:introduction}

Gravitational waves (GWs), first predicted by Einstein's theory of General Relativity \cite{EinsteinRN2844}, have been confirmed by LIGO and Virgo collaborations \cite{LIGO2016RN2843, LIGOGW170817RN2846}. 
These waves can carry unique information about their cosmic origins, providing a new way to explore the universe \cite{Schutz2009RN2905, Maggiore2007RN2848, Cai:2016hqj, ShuJin2024RN3155,LijingMengyao2024RN3154,WangShaoJiang2024RN3156}, 
and thus their detection has attracted many attentions
\Authorfootnote 
in various observational window of GW astronomy \cite{NANOGrav2023RN3109, PPTA2023RN3110, EPTA2023RN3111, CPTA2023RN3112, Amaro-SeoaneRN579LISA, Ade2016RN2863,KawamuraRN3116DECIGO,LuoJun2016RN3020,HuRN3117Tianqin,WangHanRN3118Taiji,Taiji2021RN3152,Lijing2023RN3153,LijingMengyao2024RN3154}. 
The Stochastic Gravitational Wave Background (SGWB) is expected to contain contributions 
from a variety of astrophysical and cosmological sources, including binary black holes and neutron stars \cite{Abbott2016RN2851, LIGOGW170817RN2846}, relic gravitational waves (RGWs) \cite{Grishchuk1975, Starobinsky1979, YangZhang2005, CaiWang2021}, early universe phase transitions \cite{Kamionkowski1997RN2853, Maggiore2007RN2848, Huang:2016odd}, cosmic strings \cite{Vachaspati1985RN2854, Vilenkin1994RN2855}, and other quantum processes in the early universe \cite{Ashtekar2011RN3125,Brandenberger2017RN3126,Cai2012RN3124,Novikov2024RN3122},
and so on. 
Detecting the SGWB can provide valuable insights into those phenomena and potentially open a new window into the early universe \cite{Kamionkowski1997RN2853, Maggiore2007RN2848}.

Currently, there have been many existing methods for detecting GWs, such as ground-based laser interferometers \cite{LIGO2016RN2843, Acernese2014RN2859}, pulsar timing arrays \cite{NANOGrav2023RN3109, PPTA2023RN3110, EPTA2023RN3111, CPTA2023RN3112, CaiHe2023pta}, and cosmic microwave background (CMB) polarization measurements \cite{Ade2016RN2863}, and each method has its own features. 
For instance, ground-based detectors are sensitive to high-frequency gravitational waves and less effective at detecting low-frequency waves \cite{LIGO2016RN2843, Acernese2014RN2859}. 
Pulsar timing arrays can detect very low frequency waves but require long observation times and rely on the number and distribution of suitable pulsars \cite{Manchester2013RN2866, Hobbs2013RN2861}. 
Optical Lattice Clocks (OLCs) detectors, which trap atoms in an optical lattice and use the transition frequency of the atoms as a reference \cite{Katori2003RN2868, Takamoto2005RN2882}, offer unprecedented precision and stability \cite{Hinkley2013RN2872, Bloom2014RN2873,JunYe2024OLCprecise}, and have been used for a variety of precision measurements, including tests of fundamental physics \cite{Rosenband2008RN2885, Blatt2008RN2875}, timekeeping \cite{Hinkley2013RN2872, Bloom2014RN2873}, and geodesy \cite{Nicholson2015RN2878, Takano2016RN2903}, as well as detecting GWs \cite{Kolkowitz2016RN2902, TinoBassi2019RN2674, HeFeifan2020}.
Detection of GWs by OLCs leverages the fact that GWs induce minute changes in spacetime, affecting light frequency, which can be compared with OLC frequencies to extract GW signals. 

Instead of the aforementioned mechanism that focused on detecting GWs from a fixed source, we in this article propose to use space-based OLCs to detect SGWB by the method of cross-correlation \cite{Allen1997, AllenRomano1999} and study the potential enhancement of sensitivity. 
The detection of SGWB presents a significant challenge due to the presence of various noise sources in the measurement process and similar statistical properties of noise and SGWB signals.
%Space-based OLC detector offers a promising approach to detect GWs due to their exceptional precision and stability \cite{Kolkowitz2016RN2902}. 
The detection of SGWB by using the OLC detector presents several scientific advantages. 
Firstly, OLCs offer extremely high precision in frequency measurements, which is crucial for detecting the minute perturbations caused by GWs. 
Secondly, the space-based nature of the OLC detector eliminates terrestrial noise sources, such as seismic and thermal noise. 
Furthermore, the configuration of OLC detectors arranged in a line along the Earth's orbit is highly stable. This stability allows for longer baseline between spacecraft, enhancing the sensitivity to detect low-frequency SGWB. 
And this setup facilitates the construction of more than two sets of detectors, enabling cross-correlation between them to further improve sensitivity.
In the context of SGWB detection, it is essential to distinguish between the GW signal and the noise. One approach to achieve this is through cross-correlation method between multiple OLCs.

We conduct a comprehensive analysis of the sensitivity curves of OLC detectors, detailing the mathematical framework within which they operate. 
The response function that translates the SGWB into detector outputs is discussed. 
By examining the statistical properties of the signal and the noise characteristics of the system.
Our analysis examines the statistical properties of the signal and the noise characteristics of the system, 
and study main noise sources that affect detector performance, such as quantum projection noise (QPN), photon shot noise (PSN), and acceleration noise (AN). 
Integrating the detector's response function with its noise characteristics, we delineate the sensitivity curves for individual OLC detectors. 
This is a critical aspect of GW detectors, characterizing their capability to detect SGWB at various frequencies and clearly representing the relationship between the minimum detectable amplitude of SGWB. 
However, the sensitivity of a single OLC detector is worse than LISA at whole.
To address this problem, we discuss the impact of cross-correlation techniques between pairs of OLC detectors, which can significantly enhance SNR and improve the overall detection capability. 
By coordinating the deployment of multiple spacecraft, the OLC system can effectively mitigate noise.

The sensitivity curve of cross-correlation reveals that OLC detectors have lower sensitivity levels compared to detectors like LISA. 
This unique characteristic highlights the potential of OLC detectors to capture signals from a variety of astrophysical sources, particularly those in the frequency range of $(10^{-4}, 1) \rm Hz$ which is broader than LISA.

The article is organized as follows. 
In Section \ref{sec:single_OLC}, we provide an overview of the OLC detector configuration and its operational principles, and discusses the sensitivity analysis of a single OLC detector, comparing its response characteristics to those of existing detectors like LISA. 
In Section \ref{sec:pair_OLCs}, we explore the advantages of employing cross-correlation techniques between two OLC detectors, highlighting the improvements in signal detection and noise suppression. 
In Section \ref{sec:instability_OLCs}, we analyze the impact of OLC instability on the sensitivity curve, evaluating how advancements in clock precision could enhance detection capabilities.
Finally, Section \ref{sec:conclusion} presents our conclusions and outlines potential future work.

\section{Sensitivity curve for SGWB with a single OLC detector}
\label{sec:single_OLC}

\begin{figure}[H]
\centering
\includegraphics[width=0.4\textwidth]{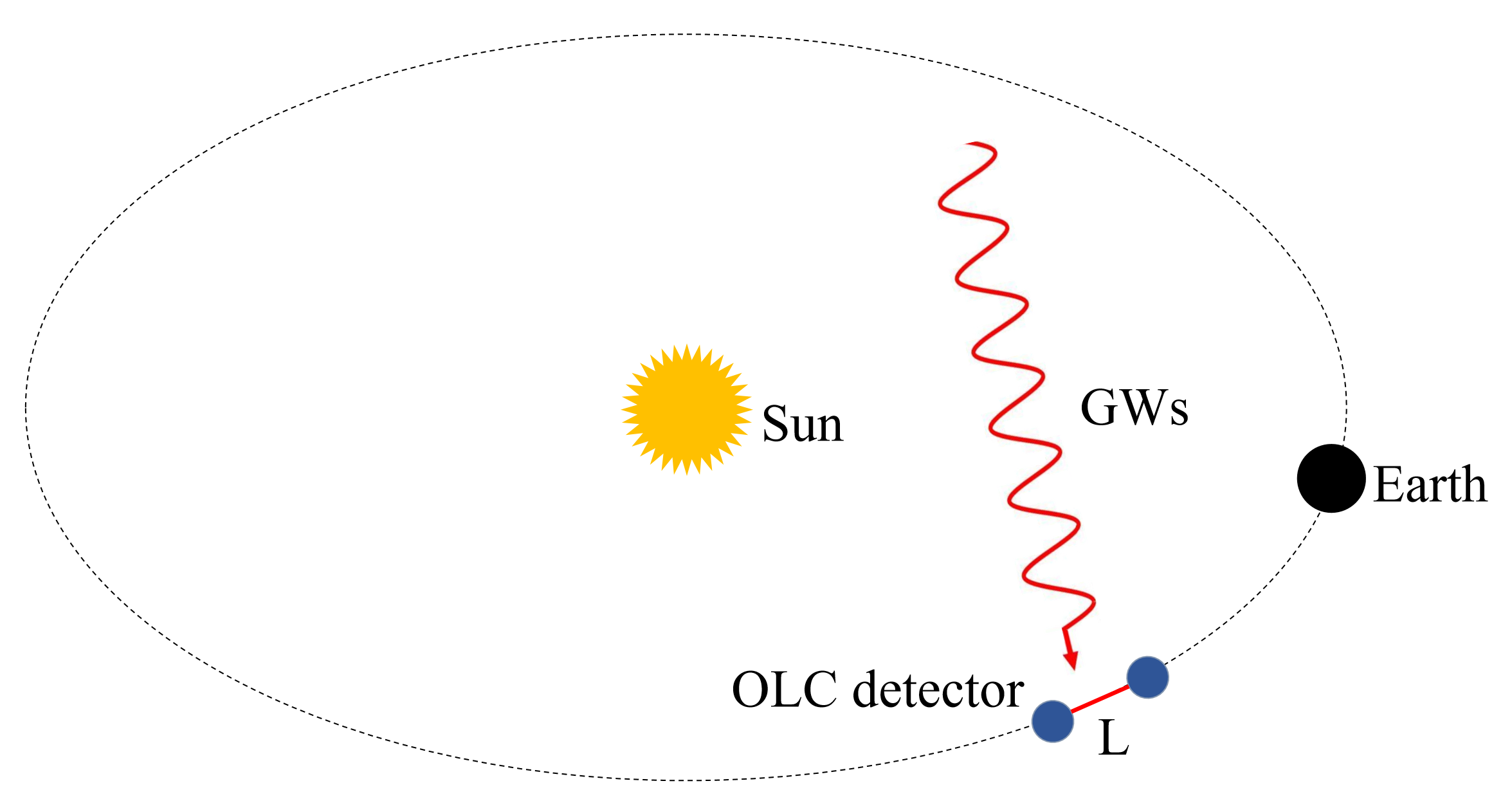}
\caption{Configuration of the OLC detector. }
\label{fig:detector_config_single}
\end{figure}

We adopt the OLC detector scheme proposed by Ref.~\cite{Kolkowitz2016RN2902} as shown in Fig.~\ref{fig:detector_config_single},  that two spacecraft, each carrying an OLC and connected by a laser link, are placed in the Earth's rotational orbit. 
The arm length between two spacecraft is $5 \times 10^{10}$\,m, which is much longer than the baseline on Earth.
Passing GWs  will induce relative motion between the two spacecraft, leading to an effective Doppler shift in the frequency of the laser between them. 
This frequency shift can be measured by the two synchronized OLCs, so to detect GWs.

GWs affect spacetime through the metric,
\[
ds^{2} = -dt^2+(\delta_{ij}+h_{ij})dx_idx_j~,
\]
where $h_{ij}$ represents the gravitational wave component. 
GWs induce a Doppler shift in the laser frequency given by
\begin{equation}\label{response_single_s}
h_o(\widehat{\Omega}, f, \mathbf{x}, t) = \frac{\Delta \nu}{\nu} = \mathbf{D}(\widehat{\Omega}, f) : \mathbf{h}(\widehat{\Omega}, f, \mathbf{x}, t) ~,
\end{equation}
with 
\begin{equation}
\mathbf{D}(\widehat{\Omega}, f) =\frac{1}{2} (\mathbf{u} \otimes \mathbf{u}) \mathcal{T}(\mathbf{u} \cdot \widehat{\Omega}, f) ~,
\end{equation}
%\begin{widetext}
\begin{equation}\label{T_f}
\mathcal{T}(\mathbf{u} \cdot \widehat{\Omega}, f) = \frac{2\pi f L}{c} \mathrm{sinc} \Big[ \pi f \frac{L}{c} (1-\mathbf{u} \cdot \widehat{\Omega}) \Big] 
e^{i \left( \frac{3}{2} \pi 
+ \pi f \frac{L}{c} (1-\mathbf{u} 
\cdot \widehat{\Omega}) \right)} ~,
\end{equation}
%\end{widetext}
where $(\mathbf{u} \otimes \mathbf{u}): \mathbf{h}\equiv u_i u_j h_{ij}$ 
and $h_o$ represents the output response of GWs captured by the detector without considering noises,
$\mathbf{D}(\widehat{\Omega}, f)$ is the detector response tensor, $\mathcal{T}(\mathbf{u} \cdot \widehat{\Omega}, f)$ is the transfer function,
$\mathbf{u}$ is a unit vector pointing from the first to the second spacecraft, 
and $\widehat{\Omega}$ is a unit vector in the direction of the GW propagation.
$\mathbf{D}(\widehat{\Omega}, f)$ depends solely on the intrinsic properties of the detector. 
By setting specific values for $\mathbf{u} \cdot \hat{\Omega} = 0$,  one can recover the results in Ref.~\cite{Kolkowitz2016RN2902}.
The factor $\frac{2 \pi f L}{c}$  in Eq.~\eqref{T_f} accounts for the frequency shift instead of the phase shift used in other references such as LISA \cite{CornishRubbo2003}, which should be considered in the noises for these two different approaches as shall be seen later.

\begin{figure}[H]
\centering
\includegraphics[width=0.4\textwidth]{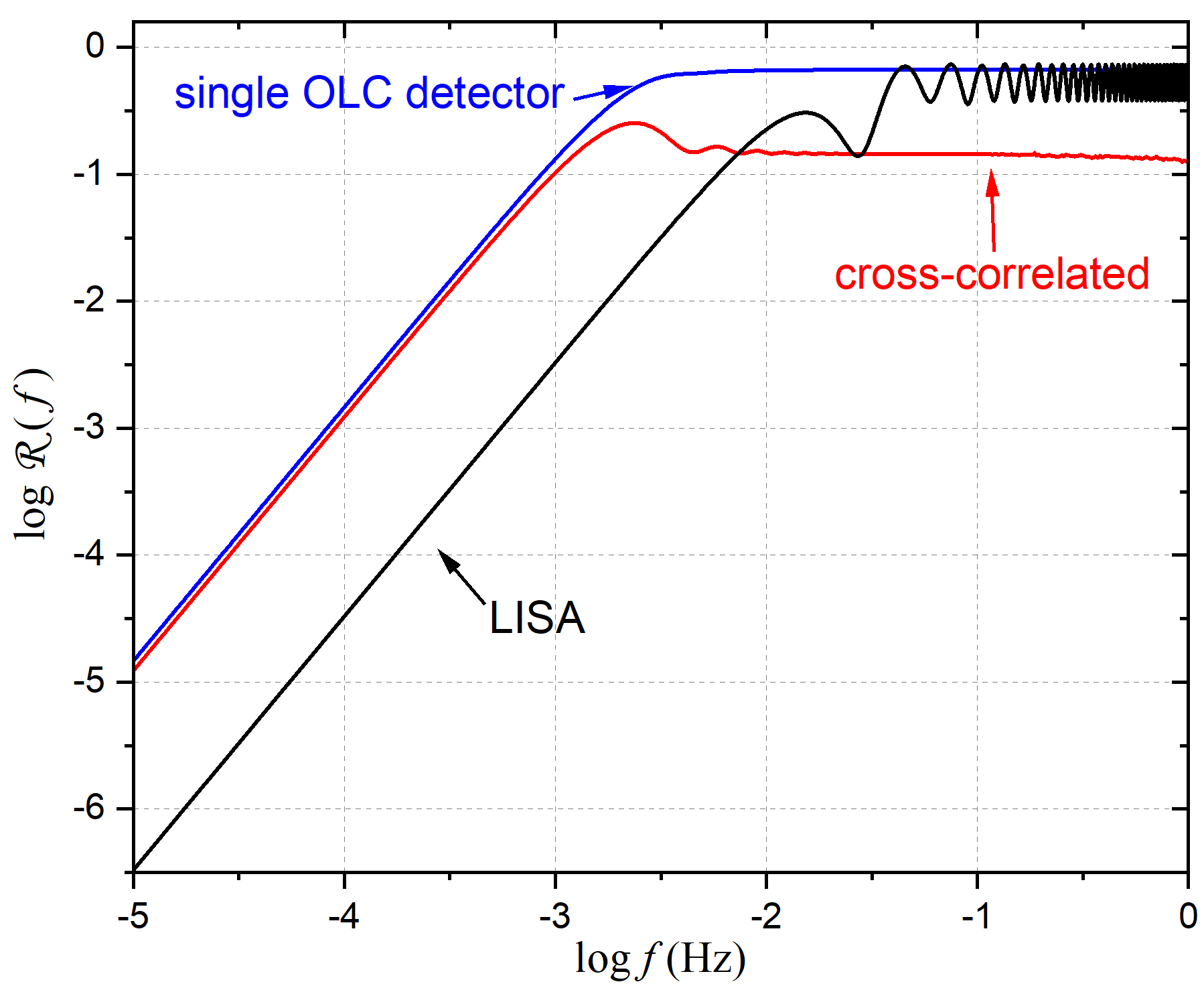}
\caption{$\mathcal{R}(f)$ for a single OLC detector (blue line), for two cross-correlated OLC detectors (red line), and for LISA (black line) as comparison.}
\label{fig:R_f_compare}
\end{figure}

Next, let us focus on the main target of this paper, SGWB. A general SGWB can be expressed in Fourier expansion as follows:
\begin{equation}\label{hijFourierExp}
h_{i j}
={\displaystyle \sum_{A}} \int_{-\infty}^{\infty} d f \int d \hat{\Omega} \,\tilde{h}_A(f, \hat{\Omega}) e^{-i2 \pi f [t-\frac{\hat{\Omega} \cdot ( \mathbf{x} - \mathbf{x}_0 )}{ c}]} \epsilon_{i j}^A(\hat{\Omega}) ~,
\end{equation}
where $\hat{\Omega}$ is the direction of a single gravitational wave (GW), $A=+,\times$ represents the polarizations of the GWs, and $\mathbf{x}_0$ is the position of the spacecraft measuring the signal.
An isotropic, stationary, and unpolarized SGWB can be taken with zero mean {\small$\langle\tilde{h}_A(f, \hat{\Omega})\rangle =0$}, and its the auto-correlation
\begin{equation}\label{h2 emsemble}
\langle
\tilde{h}_A^*(f, \hat{\Omega}) 
\tilde{h}_{A^{\prime}}(f^{\prime}, \hat{\Omega}^{\prime})\rangle  
=
\frac{1}{2} \delta\left(f-f^{\prime}\right) \frac{\delta^2(\hat{\Omega}, \hat{\Omega}^{\prime})}{4 \pi} \delta_{A A^{\prime}} S_h(f) ~,
\end{equation}
depends on the power spectral density (PSD) $S_h(f)$.
Substituting $h_{ij}$ from Eq.~\eqref{hijFourierExp} into Eq.~\eqref{response_single_s} and using the properties above, we obtain the auto-correlation of the output response as:
\begin{equation}\label{AutoCorrS}
\langle \tilde h^*_o(f)\tilde h_o(f') \rangle
=\frac{1}{2}\delta(f-f')S_h(f)\mathcal{R}(f) ~,
\end{equation}
with
\bl
&
\mathcal{R}(f) =  \int \frac{d \widehat{\Omega}}{4 \pi} \sum_A F^A(\widehat{\Omega}, f) F^{A*}(\widehat{\Omega}, f) 
\label{response_single}
\\
&
F^A(\widehat{\Omega}, f) =  \mathbf{D}(\widehat{\Omega}, f): {\bm\epsilon}^A(\widehat{\Omega})  ~.
\label{FA}
\el
The structure of Eq.~\eqref{AutoCorrS} is similar to Eq.~\eqref{response_single_s}, except that Eq.~\eqref{response_single_s} pertains to the detector's response to a single wave source, whereas Eq.~\eqref{AutoCorrS} pertains to the response to the SGWB.
$S_h(f)$ is determined by the SGWB, and the transfer function $\mathcal{R}(f)$ is determined by the design and configuration of the detector which characterizes how effectively the detector responds to GWs at different frequencies.
We plot $\mathcal{R}(f)$ of Eq.~\eqref{response_single}  as the blue line in Fig.~\ref{fig:R_f_compare}. 
For comparison, the response of LISA is also shown by the black line, which is obtained by converting the results from Ref.~\cite{CornishLarson2001} using a factor of $(2 \pi f d_{\text{L}}/c)^2$, where $d_{\text{L}} = 5 \times 10^9 \, \text{m}$ is the arm length of LISA. 
In Fig.~\ref{fig:R_f_compare}, the high-frequency response of the detector is comparable to that of LISA, but the low-frequency response is higher than LISA. 
Additionally, the transition point from low to high frequency occurs at a lower frequency of the OLC detector than LISA. 
These differences are primarily due to the longer arm length of the OLC detector compared to LISA, resulting in a lower characteristic frequency.

\begin{figure}[H]
\centering
\includegraphics[width=0.4\textwidth]{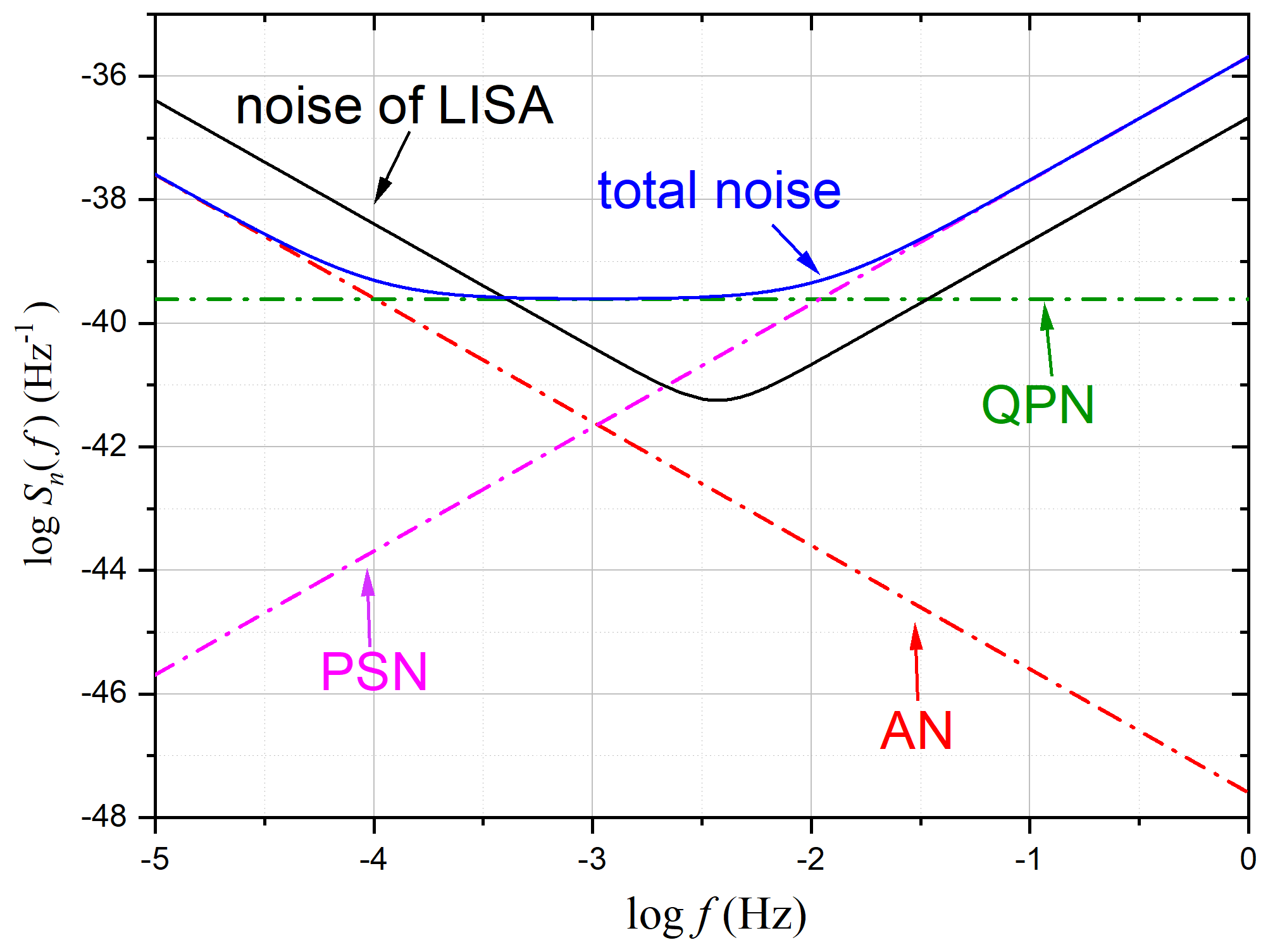}
\caption{PSDs of different noise components and the total noise. }
\label{fig:noise_all}
\end{figure}

In practical measurements,
the total signal, which includes noise unavoidably,
can be expressed as
\begin{equation} \label{total signal}
 s(t) = h_o(t) + n(t) ~,
\end{equation}
where $ n(t) $ is assumed to be Gaussian noise with zero mean, $ \langle n(t) \rangle = 0 $. 
The GW output response $ h_o(t) $ and the noise $ n(t) $ are independent, hence $ \langle h_o(t) n(t) \rangle = 0 $. 
The statistical properties of SGWB and the Gaussian noise are similar, so the PSD of the noise is defined similarly to the auto-correlation function  \eqref{h2 emsemble} as follows
\begin{equation} \label{noise ensemble}
 \langle \tilde{n}^*(f) \tilde{n}(f') \rangle = \frac{1}{2} \delta(f - f') S_n(f) ~.
\end{equation}
$S_n(f)$ characterizes the frequency-dependent behavior of the noise. 
The noise level is mainly dominated by three kinds of sources: acceleration noise (AN), quantum projection noise (QPN), and photon shot noise  (PSN) \cite{Kolkowitz2016RN2902,HeFeifan2020}, and the total noise PSD is written as a sum,
\be\label{noise emsemble}
 S_n(f) = S_{\text{AN}}(f) + S_{\text{QPN}}(f) + S_{\text{PSN}}(f) ~.
\ee
It is important to note that detectors in space are not affected by seismic noise on Earth, which gives space-based detectors the opportunity to capture low-frequency GWs.
AN for the frequency-shift scheme due to spacecraft disturbances can be modeled as \begin{equation}
S_{\text{AN}}(f) = \frac{S_a}{(2 \pi f)^2 c^2} ~, 
\end{equation}
where $ S_a = 9 \times 10^{-30} \, \text{m}^2 \, \text{s}^{-4} \, \text{Hz}^{-1} $ is the characteristic an amplitude \cite{HeFeifan2020}, 
and a factor $\frac{2 \pi f L}{c}$ for the frequency shift mentioned before has been accounted.
AN dominates the low-frequency portion of the total noise and decreases as $ \propto f^{-2} $.
QPN arises from the quantum nature of atomic state measurements, specifically due to the probabilistic outcomes when determining the state of a finite number of atoms in OLC. 
On the other hand, PSN is a consequence of the discrete nature of photons in optical systems. 
It manifests as fluctuations in the detected signal due to the random arrival times of photons.
Both kinds of noise are pivotal in the sensitivity limits of OLC detectors.
The variances of  QPN and PSN can be expressed as follows \cite{Kolkowitz2016RN2902}
\bl\label{noiseProj}
\sigma^2_{\text{QPN}} =& \frac{1}{(2 \pi \nu)^2 T t N} ~, 
\\
\label{noisepsn}
\sigma^2_{\text{PSN}} =& 
\frac{1-e^{-B t}}{2(2 \pi \nu)^2  t^2}\left(\frac{\Delta_L}{B}+\frac{h \nu}{\eta P_B} B\right) ~,
\el
where $ \nu $ is the frequency of the clock transition, $ T $ denotes the Ramsey interrogation time, $ t $ is the total measurement time, $ N $ is the number of atoms, $ \Delta_L $ is the line-width of each laser, $ \eta $ is the detector quantum efficiency, and $ P_B $ is the power from laser A received at spacecraft B.
PSD of QPN and PSN is related to its variance through $ \sigma^2 = \int_0^\infty df \, \lvert \tilde{H}(f) \rvert^2 \, S(f) $, where the noise transfer function is $ \lvert \tilde{H}(f) \rvert^2 = \text{sinc}^2(\pi f T) $ \cite{Kolkowitz2016RN2902}. 
Upon these relations, PSDs of QPN and PSN can be transferred from variance \eqref{noiseProj} and \eqref{noisepsn} as follows,
\begin{align}
 & S_{\rm QPN}(f) = \frac{2}{(2 \pi \nu)^2 N T} ~, \\
 & S_{\rm PSN}(f) = \frac{f^2 \big[ \Delta_L+\frac{h \nu B^2}{\eta P_B} \big] }{\nu^2 \big[ (2 \pi f)^2 +B^2 \big] } ~.
\end{align}
It is obvious that QPN is white noise and independent of the detection frequency $ f $, yet PSN is increasing with the square of frequency.
Ref.~\cite{Kolkowitz2016RN2902} provides a set of parameter values:
\bl \label{parameters}
&
T = 160 \, \text{s}, N = 7 \times 10^6, \nu = 430 \, \text{THz} ,  \eta=0.5,
\nn\\
&
P_B=3\,\rm pW,  \Delta_L = 30 \mathrm{mHz}, B = \sqrt{\eta P_B \Delta_L / (h \nu)}.
\el
We adopt this set of parameters to illustrate the level of noise.
PSDs of different noise components and the total noise are illustrated in Fig.~\ref{fig:noise_all}, where the LISA noise level are also plotted from Ref.~\cite{Cornish2001prd} for the Michelson interferometer.
We find that there will be low-, middle-, and high-frequency regions, respectively dominated by AN $\propto f^{-2}$, QPN $\propto f^0$, and PSN $\propto f^2$.

The frequency bands dominated by AN and PSN in the OLC detector are consistent with LISA, where the slope of the noise PSD curve of the OLC detector matches that of LISA. 
However, in the middle frequency band, the OLC detector experiences QPN dominance, differing from LISA.
This is primarily due to the increased arm length of the OLC detector
which shifts AN to lower frequencies;
and the OLC's dual-spacecraft configuration is more stable and better bounded by gravity than LISA's three spacecraft, which lowers AN level.
The shift of AN causes QPN to dominate in the middle frequency range.
As for the PSN dominated region, from the variances  \eqref{noiseProj} and \eqref{noisepsn}, it is evident that $\sigma^2_{\text{QPN}} \propto \frac{1}{t}$ and $\sigma^2_{\text{PSN}} \propto \frac{1}{t^2}$. 
As the total observation time $t$ increases, QPN becomes dominant, consistent with the description in Ref.~\cite{Kolkowitz2016RN2902}. 
However, this dominance is not in the entire frequency range.
Typically, a larger time $t$ implies a smaller frequency $f$,
so, as described above, QPN can only dominate relatively low-frequency bands, while PSN still dominates at higher frequencies. 
Therefore, the time-domain and frequency-domain descriptions, along with the corresponding two curves of QPN and PSN in Fig.~\ref{fig:noise_all}, are consistent.
Comparing the total noise of the OLC detector  with LISA, in the high frequency range, the noise level of the OLC detector is higher than that of LISA. 
This is because the Michelson interferometer of LISA  has two arms, which can reduce noise through dual-arm interference. 
However, although the OLC detector has only one arm, both detectors are on the Earth's rotational orbit, providing relative stability. 
This allows for a longer arm length, resulting in relatively lower noise in the low frequency range.

\begin{figure}[H]
    \centering
    \includegraphics[width=0.4\textwidth]{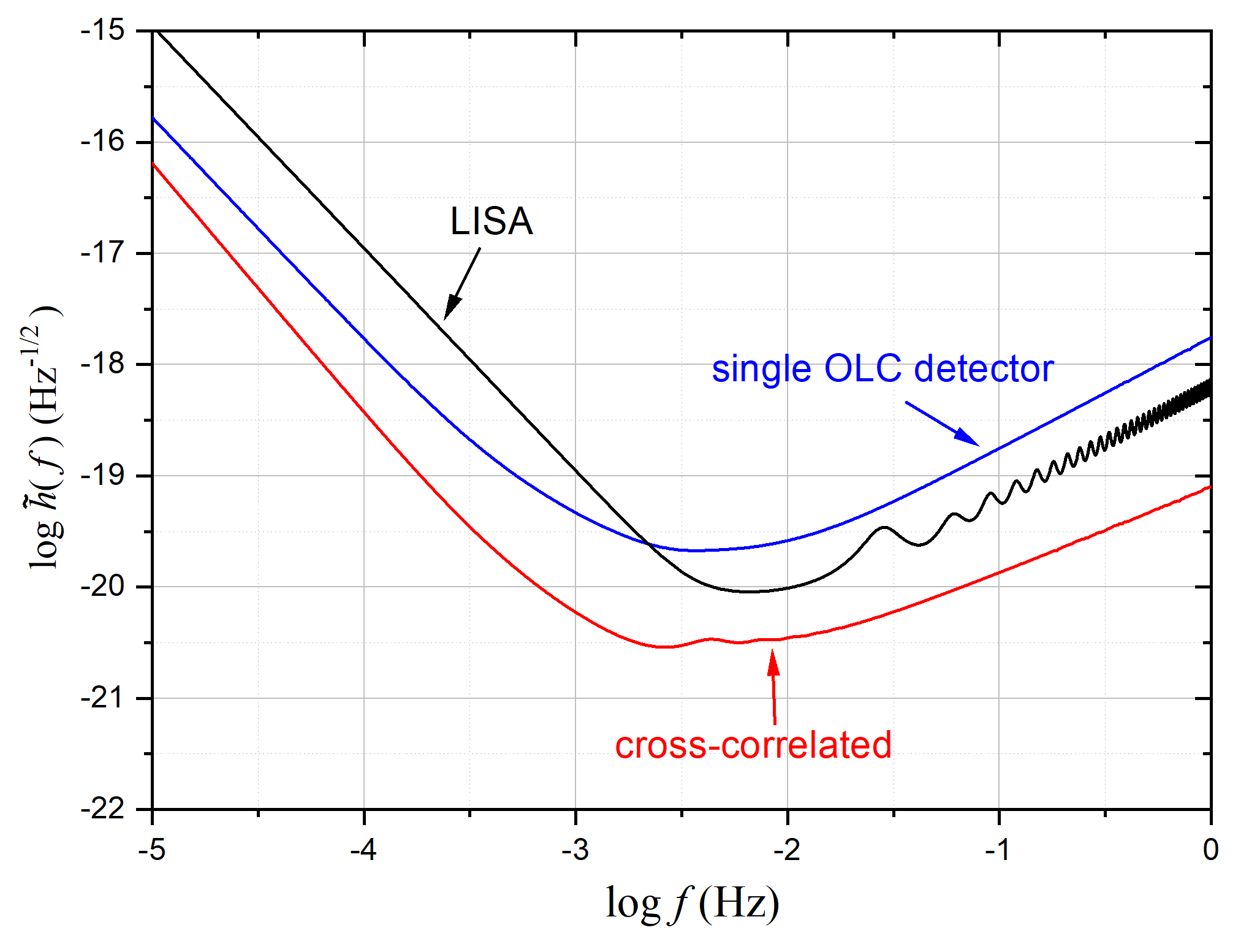}
    \caption{Sensitivity curves for a single OLC detector (blue line), a pair of cross-correlated OLC detectors (red line), and LISA (black line).}
    \label{fig:sensitivity_curve}
\end{figure}

Using the response function \eqref{response_single} and the noise \eqref{noise emsemble}, the sensitivity curve of a single OLC detector is defined as follows
\begin{equation}\label{sen_singleDEF}
\tilde{h}(f) = \sqrt{\frac{S_n(f)}{{\cal R}(f)}} ~,
\end{equation}
which characterizes the minimum detectable characteristic spectrum $ h_c(f)= \sqrt{S_h(f)}$ of SGWB by the OLC detector in the frequency domain,
and one can define \be\label{SNRsingle}
\text{SNR} = \frac{h_c(f)}{\tilde{h}(f)} ~.
\ee
The sensitivity curve takes into account both noise and the instrument's response, which is crucial for understanding the overall performance of the detector and its ability to detect SGWB.    
We plot $ \tilde{h}(f) $ in Fig.~\ref{fig:sensitivity_curve}, where the sensitivity curve for LISA \cite{Cornish2001prd} is also shown.
It can be seen that the shapes of two curves are similar. 
This similarity arises because both are primarily influenced by PSN at high frequencies and QPN at low frequencies. 
Additionally, the shapes of their $\mathcal{R}(f)$ are also similar, as shown in Fig.~\ref{fig:R_f_compare}.
There are three main differences between them. 
The first difference is that in the mid-frequency region, OLC is mainly affected by QSN, resulting in different curve shapes in the range of $10^{-4} \, \rm{Hz}$ to $10^{-2} \, \rm{Hz}$. 
The second difference is that, due to the longer arm length of an OLC detector compared to LISA, the most sensitive frequencies are not the same. 
The third one and the most important one is that the lowest point in the sensitivity curve of the OLC detector is higher than that of LISA.
In the next section, we will explore potential improvements of sensitivity by using two sets of OLC detectors.

\section{Cross-correlated detection for SGWB}
\label{sec:pair_OLCs}

\begin{figure}[H]
\centering
\includegraphics[width=0.4\textwidth]{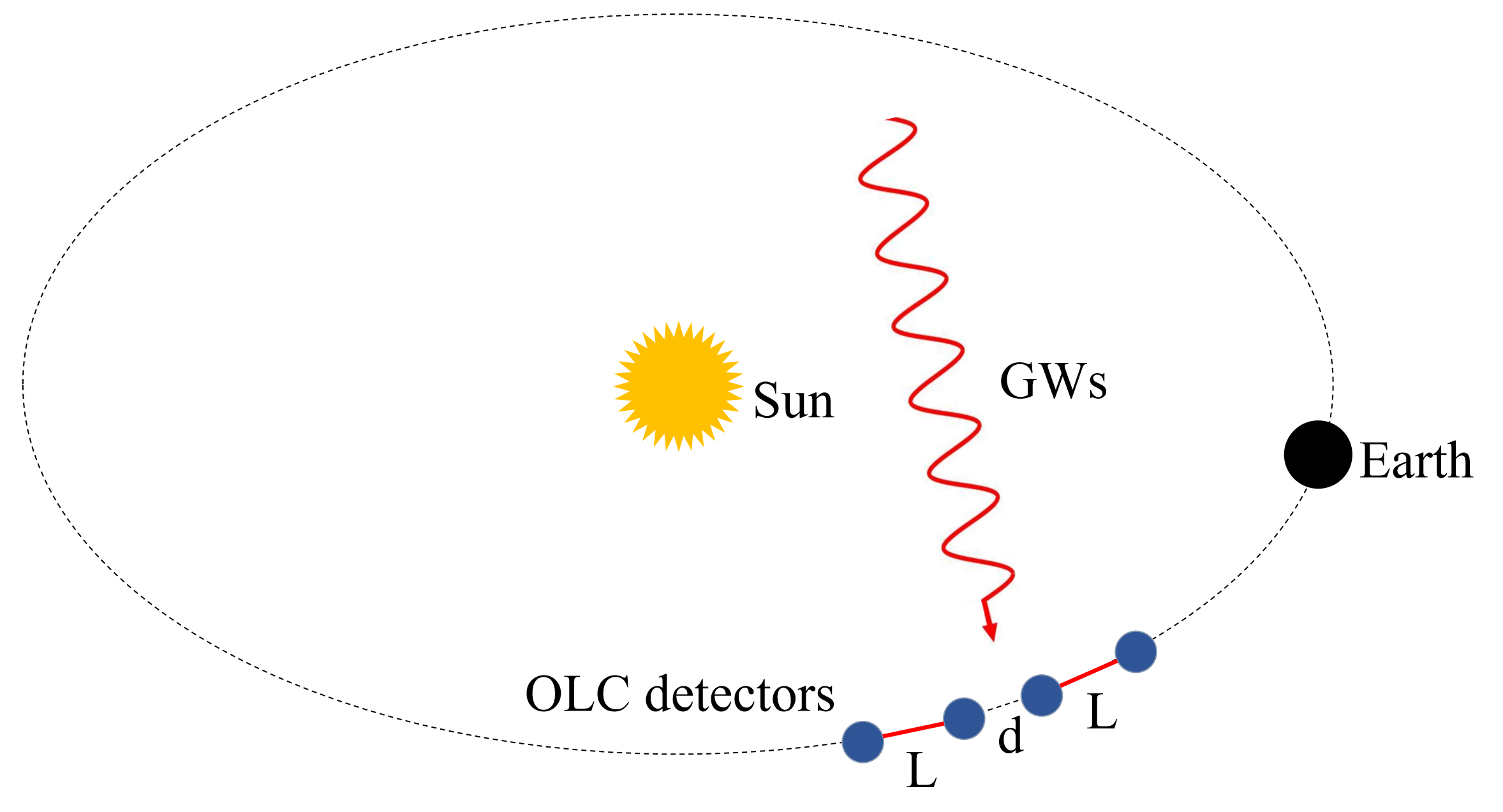}
\caption{Configuration of two sets of OLC detectors. }
\label{fig:detector_config_corr}
\end{figure}

Due to the almost identical statistical properties of the Gaussian noise in the detectors and the SGWB signal, SGWB can only be detected when it significantly exceeds the noise level.
Therefore, we need to find methods to suppress noise.
In this paper, to reduce the noise level, we adopt the cross-correlation method \cite{Allen1997,AllenRomano1999}.
This represents a potential modification to the existing OLC proposal, suggesting the deployment of four spacecraft instead of two, thus creating two independent OLC detectors and combining their outputs to achieve higher sensitivity. 
This adjustment is economically advantageous, with the total cost being significantly less than twice that of the current proposal.
This configuration is illustrated in Fig.~\ref{fig:detector_config_corr}
where four spacecraft are aligned along the Earth's rotational orbit around the Sun, forming two sets of detectors and providing high orbital stability.
The distance between two detector pairs is set to be $5 \times 10^6~\mathrm{m}$,
which is large enough that the noise in different spacecraft can be considered independent,
\begin{equation}\label{noise_n1n2}
    \langle \tilde n_1^*(f) \tilde n_2(f') \rangle = 0 ~,
\end{equation}
where the subscript $i=1,2$ labels two OLC detectors.
%the large separation between the detector pairs ensures that the noise can be treated as independent, i.e.,
%\begin{equation}
%    \langle \tilde n_1^*(f){} \tilde n_2 \left(f'\right) \rangle
%    =0.
%\end{equation}
However, the SGWB signals received by two detectors are correlated,
\begin{equation}\label{signal_h1h2}
\langle \tilde h_1^*(f) \tilde h_2(f') \rangle 
= \frac{1}{2} \delta(f-f') S_h(f) \mathcal{R}_{12}(f) ~,
\end{equation}
where
\begin{equation}\label{R12corr}
    \mathcal{R}_{12}(f) 
    =  
    {\displaystyle \sum_{A}} \int \frac{d \widehat{\Omega}}{4 \pi} e^{i 2 \pi  f\frac{  \hat{\Omega} \cdot \mathbf{x}_{12} }{c}} F_1^A{}^*(\widehat{\Omega}, f) F_2^A(\widehat{\Omega}, f) ~,
\end{equation}
where $F^A_1=F^A_2=F^A$ with $A=+,\times$ is given in Eq.~\eqref{FA}.
A larger $\mathcal{R}_{12}(f)$ indicates a better capability for a pair of detectors to convert the incident SGWB into a detectable signal. 
The structure of Eq.~(\ref{R12corr}) is similar to Eq.~\eqref{response_single}, but it has an additional exponential factor determined by the relative positions of two detectors and includes a cross term.
The response function $\mathcal{R}_{12}(f)$ is depicted by the black line in Fig.~\ref{fig:R_f_compare}. For a single OLC and cross-correlation pairs, the responses are similar at low frequencies.
However, at high frequencies, the response of the cross-correlation pairs is lower than that of a single one. 
Consequently, the cross-correlation method does not enhance the ability to convert the SGWB signal into a detectable output. 
Nevertheless, as shall be seen later, the SGWB signal can grow faster than the noise over time, potentially leading to a significant increase in SNR.

The total output signal from two detectors can be expressed as $s_i(t) = h_{oi}(t) + n_i(t)$, where $i=1,2$ labels each detector. The correlation signal between the signals of two detectors can be defined as follows \cite{Allen1997,AllenRomano1999}
\begin{align}\label{cross_signal}
C = & \int_{-\tau/2}^{\tau/2} dt \int_{-\tau/2}^{\tau/2} dt' s_1(t) s_2(t') Q(t-t') 
\nn\\
= & \int_{-\infty}^{\infty} df \int_{-\infty}^{\infty} df' \delta_\tau(f-f') \tilde{s}_1^*(f) \tilde{s}_2(f') \tilde{Q}(f') ~,
\end{align}
where $\delta_\tau(f)= \int_{-\tau/2}^{\tau/2} dt \, e^{-2 \pi i f t} = \frac{\sin(\pi f \tau)}{\pi f}$ with $\delta_\tau(0) = \tau$ and  $\underset{\tau\to \infty}{\text{lim}}\delta_\tau(f)=\delta(f)$ as the Dirac delta function. 
$Q(t-t')$ is an undetermined filter function. 
By using the relations \eqref{noise_n1n2} and \eqref{signal_h1h2}, the expectation value of $C$ is calculated,
\begin{equation}\label{expec_Corr}
 \mu = \langle C \rangle 
 = \tau \int_0^\infty df S_h(f)\mathcal{R}_{12}(f)\tilde Q(f) ~,
\end{equation}
and the covariance is
\begin{equation} \label{sigma2}
\sigma^2 
= \langle C^2 \rangle - \langle C \rangle^2
= \tau \int_{0}^\infty df \, |\tilde Q(f)|^2 \, S_n^2(f) ~.
\end{equation}
In the above expression, we assume that the noise levels in two detectors are identical, i.e., $S_{n1}(f) = S_{n2}(f)$, and that the noise within a single detector dominates over the SGWB signal, i.e., $S_n(f) > S_h(f) \mathcal{R}(f)$. This condition is generally considered valid; otherwise, cross-correlation would not be necessary.
A more general expression can obtained by replacing $S_n^2(f)$ with $S_{1n}(f) S_{2n}(f) + \mathcal{R}(f) [S_{1n}(f) + S_{2n}(f)] S_h(f) + [\mathcal{R}^2(f) + \mathcal{R}_{12}^2(f)] S_h^2(f)$ \cite{CornishLarson2001},  where the assumption of noise dominance is not required.

The optimal filter function $\tilde Q(f)$ in \eqref{cross_signal} can be derived by substituting Eqs.~\eqref{expec_Corr} and \eqref{sigma2} into $\text{SNR}_{12}$ and maximizing it \cite{Allen1997}. 
The resulting optimal one is 
$\tilde Q(f) = \frac{S_h(f)\mathcal{R}_{12}(f)}{S_n^2(f)}$. Consequently, the signal-to-noise ratio is obtained as
\begin{equation}\label{SNR12}
\text{SNR}_{12} = \sqrt{2T} \left[\int_0^\infty df \frac{S_h^2(f) \mathcal{R}_{12}^2(f)}{S_n^2(f)}\right]^{1/2} ~.
\end{equation}
By comparing Eqs.~\eqref{expec_Corr}, \eqref{sigma2}, and \eqref{SNR12}, it is evident that $\mu$ accumulates linearly with time, whereas $\sigma$ accumulates as the square root of time, resulting in a $\sqrt{\tau}$ growth factor in the SNR$_{12}$. 
For plotting the sensitivity curve of a detector pair and comparing it with a single detector, we express SNR$_{12}$ as a function of frequency. By writing the frequency integral in \eqref{SNR12} over a small interval centered at $f$ with width $\Delta f$, we obtain
\begin{equation}
\text{SNR}_{12}(f) \simeq \sqrt{\tau\Delta f} \, S_h(f) \, \overline{\left(\frac{\mathcal R_{12}^2(f)}{S_n^2(f)}\right)}^{1/2} ~,
\end{equation}
where $\overline{X(f)}$ denotes the average value of $X(f)$ over this small interval.
The effective sensitivity curve of a pair of detectors is defined analogously to SNR of a single detector in \eqref{SNRsingle} as follows
\begin{equation}\label{sensi}
\tilde h_{12}(f) = \frac{h_c}{\sqrt{\text{SNR}_{12}}} 
= \frac{1}{(\tau\Delta f)^{1/4}} \overline{\left(\frac{\mathcal R_{12}^2(f)}{S_n^2(f)}\right)}^{-1/4},
\end{equation}
which depends on $\tau$ and frequency resolution $\Delta f$. 
Compared to the sensitivity of a single one \eqref{sen_singleDEF}, a key advantage of \eqref{sensi} is the improved sensitivity with increasing observation time $\tau$.
$\tilde h_{12}(f)$ is plotted as the red line in Fig.~\ref{fig:sensitivity_curve}, with $\Delta f = f/10$ and $\tau = 1$ year. 
It is noteworthy that the cross-correlated sensitivity is approximately an order of magnitude higher than that of a single detector and surpasses LISA's sensitivity across the entire frequency range. This indicates that the cross-correlation method significantly enhances the capability of OLC detectors to detect SGWB and provides a potential path for future SGWB detection from $O(10^{-4})\,{\rm Hz}$ to $O(1)\,{\rm Hz}$.

\section{Impact of OLC Instability}
\label{sec:instability_OLCs}

The instability of OLCs is a critical factor influencing the sensitivity of  such detectors. 
In this section, we investigate the impact of OLC instability on the sensitivity curve for detecting the SGWB. 
By assuming different precision levels of OLCs, we aim to assess the feasibility of the proposed detection scheme and its potential improvements as OLC technology advances.

Substituting the parameters  in \eqref{parameters} which is adopted from Ref.~\cite{Kolkowitz2016RN2902} into Eq.~\eqref{noiseProj}, we find that the instability limit can reach $\sigma_{\rm QPN} \sim 10^{-20}$ for $t = 1\,\mathrm{s}$. 
The current state-of-the-art OLCs exhibit fractional frequency instability at the level of $\sim10^{-19}$ \cite{JunYe2024OLCprecise}. 
This suggests that the OLC instability adopted in our analysis is the projected advancements in OLC technology in the near future.

\begin{figure}[H]
    \centering
    \includegraphics[width=0.4\textwidth]{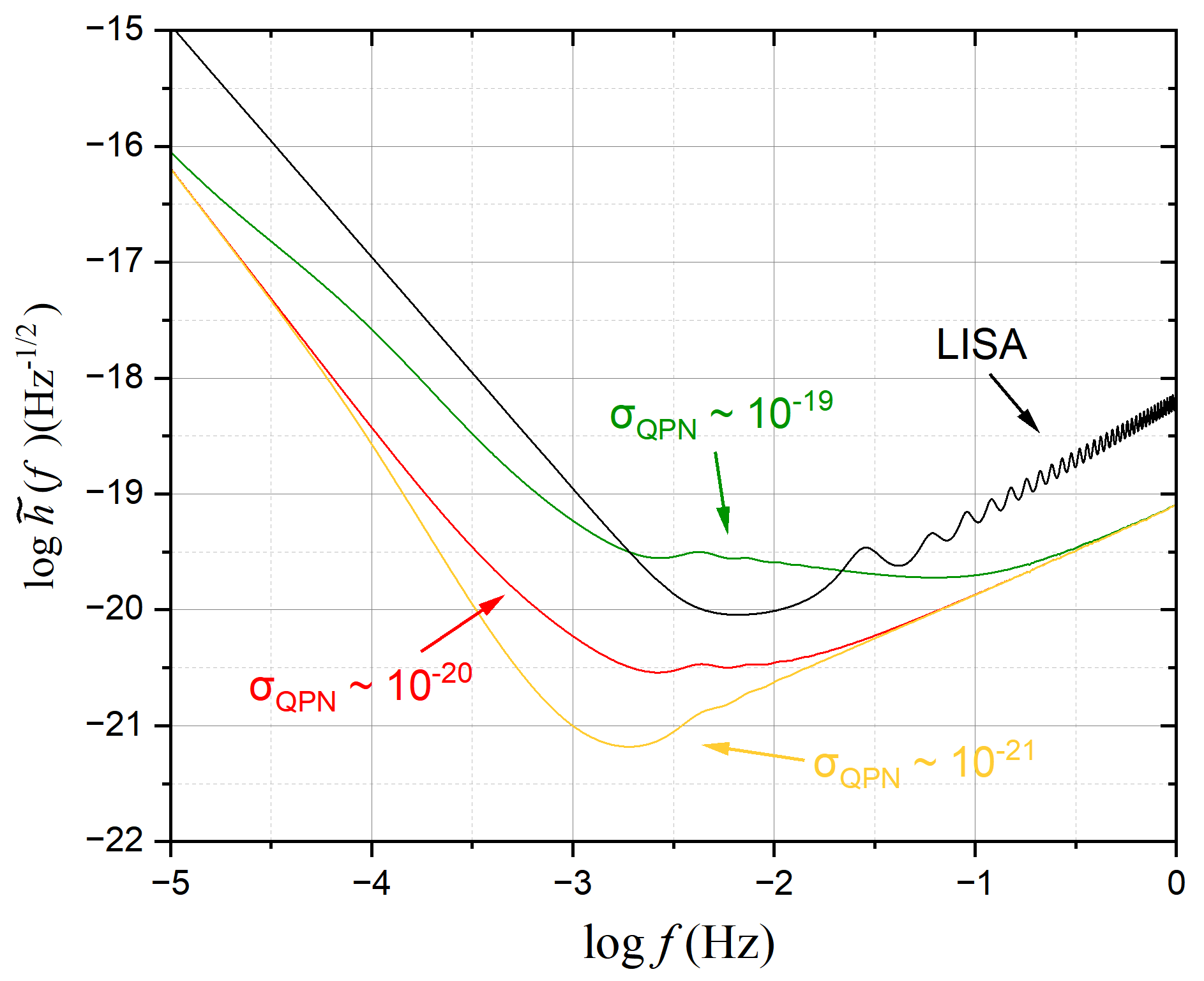}
    \caption{Sensitivity curves of cross-correlated OLC detectors under different OLC instability levels.
    }
    \label{fig:sensi_diff_sigma}
\end{figure}

To quantify the impact of OLC instability, we plot sensitivity curves under three scenarios: (1) current OLC precision ($\sim10^{-19}$), (2) precision proposed in Ref.~\cite{Kolkowitz2016RN2902} ($\sim10^{-20}$), and (3) more improved precision ($\sim10^{-21}$).
These scenarios account for both technological advancements and potential challenges in maintaining clock stability in a space-based configuration. 
The sensitivity curves for the cross-correlated OLC detectors are compared to LISA's sensitivity curve in Fig.~\ref{fig:sensi_diff_sigma},
which shows that the sensitivity improves significantly with better OLC precision. 
At the current precision of $\sigma_{\rm QPN} \sim 10^{-19}$, the sensitivity of OLC detectors around the critical frequency of $10^{-2}$ Hz does not surpass that of LISA. 
This is primarily due to the heightened QPN at this intermediate frequency range, where the instability of the OLCs dampens their sensitivity to gravitational waves.
However, a notable improvement is observed at higher frequencies, in the range of $(0.1, 1)\,\rm Hz$. 
Here, the OLC detectors exhibit enhanced sensitivity compared to LISA, offering a complementary advantage in detecting GW signals from a variety of astrophysical sources like intermediate-mass binary black hole (IMBBHs) \cite{Badurina2020RN3150}.
When the instability is further reduced to $\sigma_{\rm QPN} \sim 10^{-21}$, the sensitivity curve becomes better substantially, particularly around $10^{-3}$ Hz. 
This unprecedented sensitivity in the low-frequency region could revolutionize our understanding of ultra-long wavelength gravitational waves, potentially unveiling phenomena such as relic gravitational waves (RGWs) from the early universe \cite{ZhangYang2006RN32}, which remain elusive with current detection capabilities.

\section{Conclusion and discussions}
\label{sec:conclusion}

In this work, 
%based on the configuration proposed in Ref.~\cite{Kolkowitz2016RN2902}, 
we have conducted a comprehensive analysis of space-based OLC detectors as a promising method for space-based detection of SGWB. 
Detecting SGWB is one of the leading scientific goals in modern astrophysics, offering insights into the evolution of the early universe, and the overall distribution of binary black holes and binary pulsars in the universe, etc.
The arm lengths of space-based OLC detectors are significantly larger than Earth-based baselines, and they are not affected by terrestrial seismic noise, enabling them to capture low-frequency  GWs in the range of $(10^{-4}, 1) \, \rm Hz$.

The OLC detection scheme is based on the fundamental principle that GWs induce relative motion between two spacecraft, which can be measured through the effective Doppler shift of the laser beam frequency connecting them and the frequency comparison with the OLC.
The ability of OLCs to measure these shifts stems from its stability and precision as a measurement standard. 
We have calculated the  the OLC detector's transformation of the SGWB signal into an output signal, and quantified it by the response function $R(f)$, then plotted it as a function of frequency in Fig.~\ref{fig:R_f_compare}. We find that the response of the OLC detector is similar to that of LISA, but due to the longer arm length, its frequency turnover point is lower, giving it an advantage in detecting lower frequencies.
Our analysis considers the dominant noise sources, including AN, QPN, and PSN. 
Due to differences in arm length and configuration, the overall noise of the OLC detector differs from the Michelson interferometer of LISA, with lower noise at low frequencies but higher noise at mid to high frequencies.

Combining response and noise, we define the sensitivity curve of a single OLC detector according to Eq.~\eqref{sen_singleDEF} and plot it as a function of frequency in Fig.~\ref{fig:sensitivity_curve}. 
Our analyses reveal that, a single OLC detector has poorer detection capabilities at high frequencies compared to LISA and slightly better at low frequencies, but the advantage over LISA is not much pronounced. 
Accordingly, we explored the potential of using a pair of OLC detectors for cross-correlation to achieve SGWB detection as shown in Fig.~\ref{fig:detector_config_corr}, which is economical and with a stable orbit.
Our analyses indicate that the effective sensitivity curve of a pair of cross-correlated OLC detectors is about an order of magnitude higher than that of a single detector (see Fig.~\ref{fig:sensitivity_curve}).
This improvement in sensitivity is particularly evident across the entire frequency range, making the cross-correlation method a powerful tool for detecting SGWB.

The instability of OLCs remains a critical challenge for achieving optimal sensitivity. 
While current OLC precision levels $\sim10^{-19}$ are sufficient for initial implementations, further advancements in frequency stability to $\sim10^{-20}$  or better will be essential for fully realizing the potential of OLC-based GW detection. 
Continued progress in clock technology, including improved interrogation schemes and noise reduction techniques, will be vital to overcome this limitation and expand the detection capabilities of future OLC systems.

We also investigated the impact of OLC instability on the sensitivity of the proposed detection scheme. 
Our analysis reveals that the current state-of-the-art OLC precision, with fractional frequency instability at the level of $\sim10^{-19}$, surpasses LISA around  $(0.1, 1) \, \rm Hz$. 
Further improvements in OLC stability, reaching levels of $\sim10^{-20}$ or $\sim10^{-21}$, would dramatically enhance the sensitivity across the entire frequency range, particularly in the low-frequency regime $(10^{-3}, 10^{-2}) \, \rm Hz$. 
Such advancements could open up new opportunities for detecting more GW sources which provides a unique window into unexplored aspects of fundamental physics and cosmology.

Finally, we comment that the ability to surpass established detectors such as LISA further emphasizes the potential of the OLC detectors in GW astronomy. 
However, challenges remain in implementing the proposed method, such as mitigating noise sources, finding an optimal configuration for cross-correlation, developing signal extraction techniques or advanced data processing methods to suppress PSN, etc. 
These issues ought to be addressed in the forthcoming study such as using dynamical decoupling methods \cite{Kolkowitz2016RN2902} on cross-correlation for SGWB detection, or applying entangled quantum states \cite{Bohnet2014spinSqueeze} to achieve lower noise level of OLCs,
so that the technical prospects of space-based OLCs as gravitational wave detectors can be further consolidated.

\section*{Acknowledgements}
We are grateful to Yu-Ao Chen, Wen Zhao, Han-Ning Dai, Yuan Cao, Xibo Zhang, Yun Kau Lau and Dongdong Zhang for stimulating discussions. 
This work is supported in part by National Key R\&D Program of China (2024YFC2207500, 2021YFC2203100), CAS Young Interdisciplinary Innovation Team (JCTD-2022-20), by Fundamental Research Funds for Central Universities, by NSFC (92476203, 12433002, 12261131497),  111 Project (B23042), by CSC Innovation Talent Funds, by USTC Fellowships for International Cooperation, and by USTC Research Funds of the Double First-Class Initiative.

\section*{Conflict of interest}
The authors declare that they have no conflict of interest.

\section*{References}
\bibliographystyle{unsrt}
\bibliography{OLC_ref}

\end{multicols}
\end{document}